\documentclass{amsart}
\usepackage[cp1251]{inputenc}
\usepackage[english]{babel}
\usepackage{amsmath}
\usepackage{amssymb}
\usepackage{amsfonts}

\newtheorem{thm}{Theorem}
\begin{document}

\thispagestyle{empty}

 \title[Matrix Fourier transform  ]{ Matrix Fourier transform in dynamic theory of elasticity of
piecewise homogeneous medium
 }%

\author{O. Yaremko, E. Mogileva}%Указываем авторов

\address{Oleg Yaremko,Elena Mogileva
\newline\hphantom{iii}Penza State University,% Место работы
\newline\hphantom{iii}str. Lermontov, 37, % Адрес (улица, дом, строение и т.п.)
\newline\hphantom{iii} 440038, Penza, Russia}%  Адрес (почтовый индекс, город, страна)
\email{yaremki@mail.ru}% Ваш электронный адрес для переписки

\maketitle {\small

\begin{quote}
\noindent{\bf Abstract. } The analytical solving dynamic problems of elasticity theory for piecewise
homogeneous half-space is found. The explicit construction of direct and
inverse Fourier's vector transform with discontinuous coefficients is
presented. The technique of applying Fourier's vector transform with
discontinuous coefficients for solving problems of mathematical physics in
the heterogeneous environments is developed on an example of the dynamic
problems of the elasticity theory.
\medskip

\end{quote} }


{Penza state university, PO box 440026, Penza, Lermontov's street, 37, Russia}

\section{Introduction}

The purpose of the mathematical theory of elasticity is to define the
tension and deformations on border and inside the elastic body any form
under all load conditions. Required values are functions of coordinates and
time in dynamic problems of the theory of elasticity. Problems about
oscillations of constructions and buildings are problems of this type of
dynamic problems. Forms of oscillations and their possible changes,
amplitudes of oscillations and their increase or decrease in the course of
time, resonance modes, dynamic tension, methods of excitation and extinguish
of oscillation and others, and also problems about distribution of elastic
waves; seismic waves, and their influence on constructions and buildings,
waves arising at explosions and blows, thermoelastic waves etc. are defined
in the given problems.

Different representations of the solutions of the equilibrium equation
through functions of tension are used when solving problems by the variable
separation method. The required problem is taken to the solution of
differential equations of a more simple structure with the help of such
representations. Each functions of tension in these equations "is not
fastened" with others, but it enters into boundary conditions together with
the others. A.F.Ulitko [7] has offered rather effective method of research
of problems of mathematical physics - a method Eigen vector-valued
functions. This method is the vector analogue of the Fourier method.

The method of integral transformations is also an analytical method of the
decision of solution of problems theory of elasticity. The method of
integral transformations we consider and develop in this article. we come to
the most simple problem in space of images with the help of the integral
transformations (Fourier, Laplace, Hankel, etc.). The finding of the formula
of direct is the main difficulty in solving of problems of this approach.
Extensive enough bibliography of works on use of this method in problems of
the theory of elasticity is resulted in J.S.Ufljand's monography [2].

Problems of the theory of elasticity for heterogeneous bodies are of great
practical interest. Lame coefficients are not constant in these problems.
They are the functions of coordinates defining the field of elastic
properties of bodies. Application of analytical methods is connected with
considerable mathematical difficulties because there is no corresponding
mathematical apparatus, when the tension-strain state of bodies of the
complex configuration is researched.

Method of the vector integral transforms of Fourier is equivalent the method
Eigen vector-valued functions, however, unlike the last it can to be applied
successfully be used, applied to the solution of problems of the theory of
elasticity in a piece-wise homogeneous medium. The theory of integral
transforms of Fourier with piece-wise constant coefficients in a scalar case
was studied by Ufljand J.S. [16], [17], Najda L.S. [11], Protsenko V. S
[12], [13], Lenjuk M. P [8], [9], [10]. The vector variant of a method
adapted for the solution of problems in piece-wise homogeneous medium is
developed by the author in [2], [19]. Unknown tension in the boundary
conditions and in the internal conditions of conjugation don't commit
splitting in a considered dynamic problem, so the application of the scalar
integral transforms of Fourier with piece-wise constant coefficients does
not lead to success. Method of the vector integral transforms of Fourier
with discontinuous coefficients is used for its solution in the present
work. Conformable theoretical bases of a method are presented in item 4 for
granted. The necessary proofs by the method of contour under the scheme
developed in [2] and [19]. The closed form solution of the dynamic problem
found in the use of this method in item 4.

\section{Problem statement}

Let's consider a problem about distribution of tension in an n+1-layer
elastic semi-infinite solid $I_n^+ \times R=\left\{ {\left( {x,y}
\right):x\in I_n^+ ,y\in R} \right\}$, where $I_n^+ =\mathop \cup
\limits_{i=1}^{n+1} \left( {l_{i-1} ,l_i } \right)$. The vector of
displacement $\overline u _i $ has components $u_i ,v_i ,0$ in the case of
plane strain. If introduced two functions tension $\varphi _i \left( {x,y,t}
\right)$ and$\psi _i \left( {x,y,t} \right)$, under the condition [14],
functions are defined by the relations

\begin{equation}u_i =\frac{\partial \varphi _i }{\partial x}+\frac{\partial \psi _i
}{\partial y},
v_i =\frac{\partial \varphi _i }{\partial y}-\frac{\partial \psi _i
}{\partial x},
\end{equation}

than expressions for the component of pressure become [14]
\begin{equation}
\sigma _{ix} =\lambda _i \Delta \varphi _i +2\mu _i \left( {\frac{\partial
^2\varphi _i }{\partial x^2}+\frac{\partial ^2\psi _i }{\partial x\partial
y}} \right),
\end{equation}
\[
\sigma _{iy} =\lambda _i \Delta \varphi _i +2\mu _i \left( {\frac{\partial
^2\varphi _i }{\partial y^2}-\frac{\partial ^2\psi _i }{\partial x\partial
y}} \right)
\]
\[\tau _{ixy} =\mu _i \left( {2\frac{\partial ^2\varphi _i }{\partial
x\partial y}-\frac{\partial ^2\psi _i }{\partial x^2}+\frac{\partial ^2\psi
_i }{\partial y^2}} \right),\],

where $\lambda _i ,\mu _i $-elastic Lame constants. If to choose functions
of tension $\varphi _i $and $\psi _i $ in the form of solutions of a system
of wave equations
\begin{equation}
\frac{\partial ^2\varphi _i }{\partial t^2}=c_{1i}^2 \Delta \varphi _i ,
\quad
\frac{\partial ^2\psi _i }{\partial t^2}=c_{2i}^2 \Delta \psi _i
\end{equation}
\[
\label{eq1}
t>0,-\infty <y<\infty ,l_{i-1} <x<l_i
\]
with zero initial conditions
\begin{equation}
\label{eq2}
\varphi _i \left( {x,y,0} \right)=0,\quad \psi _i \left( {x,y,0}
\right)=0,\quad \frac{\partial \varphi _i \left( {x,y,0} \right)}{\partial
t}=0,\quad \frac{\partial \psi _i \left( {x,y,0} \right)}{\partial t}=0
\end{equation}
than the movement equations will be satisfied. The tension$p\left( {y,t}
\right)$, changing with time, is applied on the border of the body. If
tangent tension is equal to zero, than the boundary conditions become

\begin{equation}
\sigma _{1x} =-p\left( {y,t} \right),\quad \tau _{1xy} =0 \quad as \quad x=0.
\end{equation}
Let the components of the vector of displacement $\overline u _i $ and the
components of the tension tensor $\sigma _{ix} ,\quad \tau _{ixy} $ be
continuous, we get internal boundary conditions, so-called conjugation
conditions [5]:
\begin{equation}
\label{eq3}
u_i =u_{i+1} ,\quad v_i =v_{i+1} ,\quad \sigma _{ix} =\sigma _{i+1x} ,\quad
\tau _{ixy} =\tau _{i+1xy} ,\quad x=l_i ,\quad i=1,...,n.
\end{equation}
\section{Vector Fourier transform with discontinuous coefficients}

Let's develop the method of vector Fourier transform for the solution this
problem. Let's consider Sturm--Liouville vector theory [1] about a design
bounded on the set of non-trivial solution of separate simultaneous ordinary
differential equations with constant matrix coefficients
\begin{equation}
\label{eq4}
\left( {A_m^2 \frac{d^2}{dx^2}+\lambda ^2{\rm E}+\Gamma _m^2 } \right)y_m
=0,\;\,q_m^2 =\lambda ^2{\rm E}+\Gamma _m^2 ,\;\,m=\overline {1,n+1}
\end{equation}
on the boundary conditions.
\begin{equation}
\label{eq5}
\left. {\left( {\left( {\alpha _{11}^0 +\lambda ^2\delta _{11}^0 }
\right)\frac{d}{dx}+\left( {\beta _{11}^0 +\lambda ^2\gamma _{11}^0 }
\right)} \right)y_1 } \right|_{x=l_0 } =0,\quad \left. {\left\| {y_{n+1} }
\right\|{\kern 1pt}{\kern 1pt}} \right|_{x=\infty } \,<\,\infty
\end{equation}
and conditions of the contact in the points of conjugation of intervals
\begin{equation}
\label{eq6}
\left( {\left( {\alpha _{j1}^k +\lambda ^2\delta _{j1}^k }
\right)\frac{d}{dx}+\left( {\beta _{j1}^k +\lambda ^2\gamma _{j1}^k }
\right)} \right)y_k =\left( {\left( {\alpha _{j2}^k +\lambda ^2\delta
_{j2}^k } \right)\frac{d}{dx}+} \right.\left. {\left( {\beta _{j2}^k
+\lambda ^2\gamma _{j2}^k } \right) } \right)y_{k+1} ,
\end{equation}
$x=l_k ,\;\,k=\overline {1,n} ,\;\,j=1,2.,$ where
\[
y_m \left( {x,\lambda } \right)=\left( {{\begin{array}{*{20}c}
 {y_{1m} \left( {x,\lambda } \right)} \hfill \\
 \vdots \hfill \\
 {y_{rm} \left( {x,\lambda } \right)} \hfill \\
\end{array} }} \right),
\left\| {y_m } \right\|=\sqrt {y_{1m}^2 +...+y_{rm}^2 } ,m=\overline {1,n+1}
.
\]
Let for some $\lambda $ the considered the boundary problem has a
non-trivial solution
\[
y\left( {x,\lambda } \right)=\sum\limits_{k=1}^n {\theta \left( {x-l_{k-1} }
\right)\,\theta \left( {l_k -x} \right)\,y_k \left( {x,\lambda }
\right)\,+\,\theta \left( {x-l_n } \right)\,y_{n+1} \left( {x,\lambda }
\right)} .
\]
The number $\lambda $ is called an Eigen value in this case, and the
corresponding decision $y\left( {x,\lambda } \right)$ is called Eigen
vector-valued function.

$$\alpha _{11}^0 ,\beta _{11}^0 ,\gamma _{11}^0 ,\delta _{11}^0 ,\alpha
_{j1}^k ,\beta _{j1}^k ,\gamma _{j1}^k ,\delta _{j1}^k ,\alpha _{j2}^k
,\beta _{j2}^k ,\gamma _{j2}^k ,\delta _{j2}^k ,A_j -\;\left(
{j=1,2;\;\,m=1,n+1;\;\,k=1,n} \right)$$
are matrixes of the size $r\times r$.
We will required invertible
\begin{equation}
\label{eq7}
\det \;\;M_{mk} \ne 0,\;\;\lambda \in \left. {\left[ {0,\infty } \right.}
\right)
\end{equation}
for matrixes
\[
M_{mk} \equiv \left( {{\begin{array}{*{20}c}
 {\beta _{1m}^k +\lambda ^2\gamma _{1m}^k } \hfill & {\alpha _{1m}^k
+\lambda ^2\delta _{1m}^k } \hfill \\
 {\beta _{2m}^k +\lambda ^2\gamma _{2m}^k } \hfill & {\alpha _{2m}^k
+\lambda ^2\delta _{2m}^k } \hfill \\
\end{array} }} \right),\;\,m=1,2;\;\,k=\overline {1,n} .
\]
Matrixes $A_m^2 $ and $\Gamma _m^2 $ , are is $m=\overline {1,n+1} $
-positive-defined [6]. We denote
\[
\Phi _{n+1} \left( x \right)=e^{q_{n+1} xi};\;\,\Psi _{n+1} \left( x
\right)=e^{-q_{n+1} xi};\;\,q_{n+1}^2 =A_{n+1}^{-2} \left( {\lambda ^2{\rm
E}+\Gamma ^2} \right).
\]
Define the induction relations the others n-pairs a matrix-importance
functions $\left( {\Phi _k ,\Psi _k } \right),\;\;k=1,n:$
\[
\left[ {\left( {\alpha _{j1}^k +\lambda ^2\delta _{j1}^k }
\right)\frac{d}{dx}+\left( {\beta _{j1}^k +\lambda ^2\gamma _{j1}^k }
\right)} \right]\,\left( {\Phi _k ,\Psi _k } \right)=
\]
\begin{equation}
\label{eq8}
=\left[ {\left( {\alpha _{j2}^k +\lambda ^2\delta _{j2}^k }
\right)\frac{d}{dx}+\left( {\beta _{j2}^k +\lambda ^2\gamma _{j2}^k }
\right)} \right]\,\left( {\Phi _{k+1} ,\Psi _{k+1} } \right),\quad
k=\overline {1,n} ,\quad j=\overline {1,2} .
\end{equation}
Let us introduce the following notation
\[
\left. {\mathop \Phi \limits^0 _1 \left( \lambda \right)=\left[ {\left(
{\alpha _{11}^0 +\lambda ^2\delta _{11}^0 } \right)\frac{d}{dx}+\left(
{\beta _{11}^0 +\lambda ^2\gamma _{11}^0 } \right)} \right]\Phi _1 \left(
{x,\lambda } \right)\,} \right|_{x=l_0 } ,
\]
\[
\left. {\mathop \Psi \limits^0 _1 \left( \lambda \right)=\left[ {\left(
{\alpha _{11}^0 +\lambda ^2\delta _{11}^0 } \right)\frac{d}{dx}+\left(
{\beta _{11}^0 +\lambda ^2\gamma _{11}^0 } \right)} \right]\Psi _1 \left(
{x,\lambda } \right)\,} \right|_{x=l_0 } ,
\]
\[
\Omega _k =\left( {{\begin{array}{*{20}c}
 {\Phi _k } \hfill & {\Psi _k } \hfill \\
 {\Phi _k^/ } \hfill & {\Psi _k^/ } \hfill \\
\end{array} }} \right),\quad i=\overline {1,n+1} .
\]
\begin{thm}The spectrum of the problem (\ref{eq4}),(\ref{eq5}),(\ref{eq6}) is a continuous and fills
all semi axis $\left( {0,\infty } \right)$. Sturm--Liouville theory r time
is degenerate. To each Eigen value $\lambda $ corresponds to exactly $r$ linearly
independent vector-valued functions. As the last it is possible to take $r$
columns matrix-importance functions.
\[
u\left( {x,\lambda } \right)=\sum\limits_{k=1}^n {\theta \left( {x-l_{k-1} }
\right)\,\theta \left( {l_k -x} \right)\,u_k \left( {x,\lambda }
\right)\,+\,\theta \left( {x-l_n } \right)\,u_{n+1} \left( {x,\lambda }
\right)} ,
\]
\begin{equation}
\label{eq9}
u_j \left( {x,\lambda } \right)=\Phi _j \left( {x,\lambda } \right)\mathop
{\Phi _1^{-1} }\limits^{0\;\;\;} \left( \lambda \right)-\Psi _j \left(
{x,\lambda } \right)\mathop {\Psi _1^{-1} }\limits^{0\;\;\;} \left( \lambda
\right).
\end{equation}
That is
\[
y^m\left( {x,\lambda } \right)=\left( {{\begin{array}{*{20}c}
 {u_{1m} \left( {x,\lambda } \right)} \hfill \\
 \vdots \hfill \\
 {u_{rm} \left( {x,\lambda } \right)} \hfill \\
\end{array} }} \right).
\]
\end{thm}
Dual Sturm--Liouville theory consists in a finding of the non-trivial
solution of separate simultaneous ordinary differential equations with
constant matrix coefficients.
\begin{equation}
\label{eq10}
\left( {A_m^2 \frac{d^2}{dx^2}+\lambda ^2{\rm E}+\Gamma _m^2 } \right)y_m
=0,\;\,q_m^2 =\lambda ^2{\rm E}+\Gamma _m^2 ,\;\,m=\overline {1,n+1}
\end{equation}
on the boundary conditions
\begin{equation}
\label{eq11}
\left. {\left( {\frac{d}{dx}y_1^\ast \left( {\beta _{11}^0 +\lambda ^2\gamma
_{11}^0 } \right)^{-1}+y_1^\ast \left( {\alpha _{11}^0 +\lambda ^2\delta
_{11}^0 } \right)^{-1}} \right){\kern 1pt}} \right|_{x=l_0 } =0,\quad
\;\;\left\| {y_{n+1}^\ast } \right\|\,<\,\infty ,
\end{equation}
and conditions of the contact in the points of conjugation of intervals
\[
\left( {-\frac{d}{dx}y_k^\ast ,y_k^\ast } \right)\left(
{{\begin{array}{*{20}c}
 {\beta _{11}^k +\lambda ^2\gamma _{11}^k } \hfill & {\alpha _{11}^k
+\lambda ^2\delta _{11}^k } \hfill \\
 {\beta _{21}^k +\lambda ^2\gamma _{21}^k } \hfill & {\alpha _{21}^k
+\lambda ^2\delta _{21}^k } \hfill \\
\end{array} }} \right)^{-1}=
\]
\begin{equation}
\label{eq12}
=\left( {-\frac{d}{dx}y_{k+1}^\ast ,y_{k+1}^\ast } \right)\left(
{{\begin{array}{*{20}c}
 {\beta _{12}^k +\lambda ^2\gamma _{12}^k } \hfill & {\alpha _{12}^k
+\lambda ^2\delta _{12}^k } \hfill \\
 {\beta _{22}^k +\lambda ^2\gamma _{22}^k } \hfill & {\alpha _{22}^k
+\lambda ^2\delta _{22}^k } \hfill \\
\end{array} }} \right)^{-1},\quad \quad x=l_k ,\quad k=\overline {1,n} .
\end{equation}
The solution of the boundary value problem we write in the form of
\[
y^\ast \left( {\xi ,\lambda } \right)=\sum\limits_{k=2}^n {\theta \left(
{\xi -l_{k-1} } \right)\,\theta \left( {l_k -\xi } \right)\,y_k^\ast \left(
{\xi ,\lambda } \right)\,+\theta \left( {l_1 -\xi } \right)\,y_1^\ast \left(
{\xi ,\lambda } \right)+\theta \left( {\xi -l_n } \right)\,y_{n+1}^\ast
\left( {\xi ,\lambda } \right)} ,
\]
\[
y_m^\ast \left( {\xi ,\lambda } \right)=\left( {{\begin{array}{*{20}c}
 {y_{m1}^\ast \left( {\xi ,\lambda } \right)} \hfill & \cdots \hfill &
{y_{mr}^\ast \left( {\xi ,\lambda } \right)} \hfill \\
\end{array} }} \right),
\]
\[
\left\| {y_m^\ast } \right\|=\sqrt {\left( {y_{1m}^\ast }
\right)^2+...+\left( {y_{rm}^\ast } \right)^2} ,m=\overline {1,n+1} .
\]
\begin{thm}The spectrum of the problem (\ref{eq4}),(\ref{eq5}),(\ref{eq6}) is a continuous and fills
semi axis $\left( {0,\infty } \right)$. Sturm--Liouville theory r time is
degenerate. To each Eigen value $\lambda $ corresponds to exactly $r$ linearly
independent vector-valued functions. As the last it is possible to take $r$
rows matrix-importance functions.
\[
u^\ast \left( {x,\lambda } \right)=\sum\limits_{k=1}^n {\theta \left(
{x-l_{k-1} } \right)\,\theta \left( {l_k -x} \right)\,u_k^\ast \left(
{x,\lambda } \right)\,+\,\theta \left( {x-l_n } \right)\,u_{n+1}^\ast \left(
{x,\lambda } \right)} ,
\]
\[
u_j^\ast \left( {x,\beta } \right)=\left( {\mathop \Phi \limits^0 _1 \left(
\beta \right),\mathop \Psi \limits^0 _1 \left( \beta \right)}
\right)\,\Omega _j^{-1} \left( {x,\beta } \right)\left(
{{\begin{array}{*{20}c}
 0 \hfill \\
 {\rm E} \hfill \\
\end{array} }} \right)A_j^{-2} ,
\]
That is
\begin{equation}
\label{eq13}
y^{\ast j}\left( {\xi ,\lambda } \right)=\left( {{\begin{array}{*{20}c}
 {u_{j1}^\ast \left( {\xi ,\lambda } \right)} \hfill & \cdots \hfill &
{u_{jr}^\ast \left( {\xi ,\lambda } \right)} \hfill \\
\end{array} }} \right),
j=\overline {1,r} .
\end{equation}
\end{thm}
The existence of spectral functions $u\left( {x,\lambda } \right)$ and the
conjugate spectral function $u^\ast \left( {x,\lambda } \right)$ allows to
write the a vector decomposition theorem on the set of $I_n^+ $.

\begin{thm} Let the vector-valued function $f (x)$ is defined on $I_n^+ $
continuous, absolutely integrated and has the bounded total variation. Then
for any $x\in I_n^+ $ true formula of decomposition
\[
f\left( x \right)=-\frac{1}{\pi j}\int\limits_0^\infty {u\left( {x,\lambda }
\right)} \left(  \right.\int\limits_{l_0 }^\infty {u^\ast
\left( {\xi ,\lambda } \right)f\left( \xi \right)d\xi +}
\]
\[
+\left( {\gamma _{11}^0 f_1 \left( {l_0 } \right)+\delta _{11}^0 {f}'_1
\left( {l_0 } \right)} \right)+\sum\limits_{k=1}^n {\left( {\phi _1^0 \left(
\lambda \right),\psi _1^0 \left( \lambda \right)} \right)\,\Omega _k^{-1}
\left( {l_k ,\lambda } \right)\,M_{k1}^{-1} \left( \lambda \right)\cdot }
\]
\begin{equation}
\label{eq14}
\cdot \left. {\left\{ {\left( {{\begin{array}{*{20}c}
 {\gamma _{21}^k } \hfill & {\delta _{21}^k } \hfill \\
 {\gamma _{22}^k } \hfill & {\delta _{22}^k } \hfill \\
\end{array} }} \right)\,\left( {{\begin{array}{*{20}c}
 {f_{k+1} \left( {l_k } \right)} \hfill \\
 {{f}'_{k+1} \left( {l_k } \right)} \hfill \\
\end{array} }} \right)-\left( {{\begin{array}{*{20}c}
 {\gamma _{11}^k } \hfill & {\delta _{11}^k } \hfill \\
 {\gamma _{12}^k } \hfill & {\delta _{12}^k } \hfill \\
\end{array} }} \right)\,\left( {{\begin{array}{*{20}c}
 {f_k \left( {l_k } \right)} \hfill \\
 {{f}'_k \left( {l_k } \right)} \hfill \\
\end{array} }} \right)} \right\}} \right)\lambda d\lambda .
\end{equation}
The decomposition theorem allows to enter the direct and inverse matrix
integral Fourier transform on the real semi axis with conjugation points:
\[
F_{n+} \left[ f \right]\left( \lambda \right)=\int\limits_{l_0 }^\infty
{u^\ast \left( {\xi ,\lambda } \right)f\left( \xi \right)d\xi +}
\]
\[
+\left( {\gamma _{11}^0 f_1 \left( {l_0 } \right)+\delta _{11}^0 f_1^/
\left( {l_0 } \right)} \right)+\sum\limits_{k=1}^n {\left( {\phi _1^0 \left(
\lambda \right),\psi _1^0 \left( \lambda \right)} \right)\,\Omega _k^{-1}
\left( {l_k ,\lambda } \right)\,M_{k1}^{-1} \left( \lambda \right)\cdot }
\]
\begin{equation}
\label{eq15}
\cdot \left\{ {\left( {{\begin{array}{*{20}c}
 {\gamma _{21}^k } \hfill & {\delta _{21}^k } \hfill \\
 {\gamma _{22}^k } \hfill & {\delta _{22}^k } \hfill \\
\end{array} }} \right)\,\left( {{\begin{array}{*{20}c}
 {f_{k+1} \left( {l_k } \right)} \hfill \\
 {f_{k+1}^/ \left( {l_k } \right)} \hfill \\
\end{array} }} \right)-\left( {{\begin{array}{*{20}c}
 {\gamma _{11}^k } \hfill & {\delta _{11}^k } \hfill \\
 {\gamma _{12}^k } \hfill & {\delta _{12}^k } \hfill \\
\end{array} }} \right)\,\left( {{\begin{array}{*{20}c}
 {f_k \left( {l_k } \right)} \hfill \\
 {f_k^/ \left( {l_k } \right)} \hfill \\
\end{array} }} \right)} \right\}\equiv \tilde {f}\left( \lambda \right),
\end{equation}
\begin{equation}
\label{eq16}
F_{n+}^{-1} \left[ {\tilde {f}} \right]\,\left( x \right)=-\frac{1}{\pi
i}\int\limits_0^\infty {\lambda u\left( {x,\lambda } \right)\,\tilde
{f}\left( \lambda \right)d\lambda } \equiv f\left( x \right),
\end{equation}
when
\[
f\left( x \right)=\sum\limits_{k=1}^n {\theta \left( {l_k -x}
\right)\,\theta \left( {x-l_{k-1} } \right)\,f_k \left( x \right)\,+\theta
\left( {x-l_n } \right)\,f_{n+1} \left( x \right)} .
\]
\end{thm}
Let's apply the obtained integral formulas for the solution of the problem
of elasticity theory (1),(2),(\ref{eq1}),(\ref{eq2}). Let's result the basic identity of
integral transform of the differential operator
\[
B=\sum\limits_{j=1}^n {\theta \left( {x-l_{j-1} } \right)\,\theta \left(
{l_j -x} \right)\left( {A_j^2 \frac{d^2}{dx^2}+\Gamma _j^2 }
\right)\,+\theta \left( {x-l_n } \right)\left( {A_{n+1}^2
\frac{d^2}{dx^2}+\Gamma _{n+1}^2 } \right)} .
\]
\begin{thm} If vector-valued function
\[
f\left( x \right)=\sum\limits_{k=1}^n {\theta \left( {x-l_{k-1} }
\right)\,\theta \left( {l_k -x} \right)f_k \left( x \right)\,+\theta \left(
{x-l_n } \right)\,f_{n+1} \left( x \right)} ,
\]
is continuously differentiated on set three times, has the limit values
together with its derivatives up to the third order inclusive
\[
f_k^{(m)} \left( {l_{k-1} } \right)=f_k^{(m)} \left( {l_{k-1} +0}
\right),\;\,m=0,1,2,3;\quad k=\overline {1,n+1}
\]
Satisfies to the boundary condition on infinity
\[
\mathop {\lim }\limits_{x\to \infty } \;\left( {u^\ast \left( {x,\lambda }
\right)\frac{d}{dx}f\left( x \right)-\frac{d}{dx}u^\ast \left( {x,\lambda }
\right)\,f\left( x \right)} \right)=0
\]
Satisfies to homogeneous conditions of conjugation (\ref{eq6}), that basic identity
of integral transform of the differential operator $B$ hold
\[
F_{n+} \left[ {B\left( f \right)} \right]\,\left( \lambda \right)=-\lambda
^2\tilde {f}\left( \lambda \right)-\left\{ {\left( {\beta _{11}^0 f_1 \left(
{l_0 } \right)+\alpha _{11}^0 f_1^/ \left( {l_0 } \right)} \right)-}
\right.
\]
\begin{equation}
\label{eq17}
\left. {-\left( {\gamma _{11}^0 A_1^2 f_1^{//} \left( {l_0 } \right)+\delta
_{11}^0 A_1^2 f_1^{///} \left( {l_0 } \right)} \right)} \right\} .
\end{equation}
\end{thm}
The proof of theorems 1,2,3,4 is spent by a method of the method of contour
integration. Similarly presented to work of the author [19].

\section{The solution of dynamic problems of the theory of elasticity}

Let's apply on the variable $y$ Fourier transformation [4], and let's apply
on the variable $x$ the vector integral transforms of Fourier (\ref{eq15}). In the
images of Fourier series in the variable $y$the problem (1), (2), (\ref{eq1}), (\ref{eq2})
takes the form of the simultaneous equations
\[
\frac{\partial ^2\bar {\varphi }_i }{\partial t^2}=c_{1i}^2 \frac{\partial
^2\bar {\varphi }_i }{\partial x^2}-c_{1i}^2 \xi ^2\bar {\varphi }_i ,
\]
\begin{equation}
\label{eq18}
\frac{\partial ^2\bar {\psi }_i }{\partial t^2}=c_{2i}^2 \frac{\partial
^2\bar {\psi }_i }{\partial x^2}-c_{2i}^2 \xi ^2\bar {\psi }_i ,
t>0,\quad l_{i-1} <x<l_i
\end{equation}
with initial conditions
\[
\bar {\varphi }_i \left( {x,y,0} \right)=0,\quad \bar {\psi }_i \left(
{x,y,0} \right)=0,\quad
\]
\begin{equation}
\label{eq19}
\frac{\partial \bar {\varphi }_i \left( {x,y,0} \right)}{\partial t}=0,\quad
\frac{\partial \bar {\psi }_i \left( {x,y,0} \right)}{\partial t}=0
\end{equation}
where $\bar {\varphi }_i ,\bar {\psi }_i $- images of Fourier series in the
variable $y$ functions of tension
\[
\bar {\varphi }_i =\frac{1}{\sqrt {2\pi } }\int_{-\infty }^\infty {\varphi
_i } \left( {x,y,t} \right)e^{-j\xi y}dy,\quad \bar {\psi }_i
=\frac{1}{\sqrt {2\pi } }\int_{-\infty }^\infty {\psi _i } \left( {x,y,t}
\right)e^{-j\xi y}dy
\]
with boundary conditions

$\sigma _{1x} =\lambda _1 \frac{\partial ^2\bar {\varphi }_1 }{\partial
x^2}-\lambda _1 \xi ^2\bar {\varphi }_1 +2\mu _1 \left( {\frac{\partial
^2\bar {\varphi }_1 }{\partial x^2}+j\xi \frac{\partial \bar {\psi }_1
}{\partial x}} \right)=-\bar {p}\left( {\xi ,t} \right),$ as $x=0$

\begin{equation}\bar {\tau }_{1xy} =\mu _1 \left( {2j\xi \frac{\partial \bar {\varphi }_1
}{\partial x}-\frac{\partial ^2\bar {\psi }_1 }{\partial x^2}-\xi ^2\bar
{\psi }_1 } \right)=0,\quad as \quad x=0
\end{equation}
with the internal conditions of conjugation
\[
\frac{\partial \bar {\varphi }_i }{\partial x}+j\xi \bar {\psi }_i
=\frac{\partial \bar {\varphi }_{i+1} }{\partial x}+j\xi \bar {\psi }_{i+1}
,
\]
\[j\xi \bar {\varphi }_i -\frac{\partial \bar {\psi }_i }{\partial x}=j\xi
\bar {\varphi }_{i+1} -\frac{\partial \bar {\psi }_{i+1} }{\partial x}, as \quad
x=l_i
\]

\[\lambda _i \frac{\partial ^2\bar {\varphi }_i }{\partial x^2}-\lambda _i \xi
^2\bar {\varphi }_i +2\mu _i \left( {\frac{\partial ^2\bar {\varphi }_i
}{\partial x^2}+j\xi \frac{\partial \bar {\psi }_i }{\partial x}} \right)=
\]
\[=\lambda _{i+1} \frac{\partial ^2\bar {\varphi }_{i+1} }{\partial
x^2}-\lambda _{i+1} \xi ^2\bar {\varphi }_{i+1} +2\mu _{i+1} \left(
{\frac{\partial ^2\bar {\varphi }_{i+1} }{\partial x^2}+j\xi \frac{\partial
\bar {\psi }_{i+1} }{\partial x}} \right) as \quad x=l_i
\]
\[
\mu _i \left( {2j\xi \frac{\partial \bar {\varphi }_i }{\partial
x}-\frac{\partial ^2\bar {\psi }_i }{\partial x^2}-\xi ^2\bar {\psi }_i }
\right)=
\]
\begin{equation}
=\mu _{i+1} \left( {2j\xi \frac{\partial \bar {\varphi }_{i+1} }{\partial
x}-\frac{\partial ^2\bar {\psi }_{i+1} }{\partial x^2}-\xi ^2\bar {\psi
}_{i+1} } \right) as \quad x=l_i
\end{equation}

Denote $c=\mathop {\max }\limits_i \left\{ {c_{1i} ,c_{2i} } \right\}$.
Let's apply to a problem (\ref{eq18}), (\ref{eq19}), (23), (24) vector integral Fourier
transform with discontinuous coefficients, defined by formulas (\ref{eq15}) - (\ref{eq16}).
Let's put in simultaneous equations (\ref{eq4})
\[
r=2,
\quad
A_i^2 =\left( {{\begin{array}{*{20}c}
 {c_{i1}^2 } \hfill & 0 \hfill \\
 0 \hfill & {c_{i2}^2 } \hfill \\
\end{array} }} \right),\quad \Gamma _i^2 =\left( {{\begin{array}{*{20}c}
 {\left( {c^2-c_{i1}^2 } \right)\xi ^2} \hfill & 0 \hfill \\
 0 \hfill & {\left( {c^2-c_{i2}^2 } \right)\xi ^2} \hfill \\
\end{array} }} \right),
\]
in boundary conditions (\ref{eq5}) let's consider
\[
\alpha _{11}^0 =\left( {{\begin{array}{*{20}c}
 0 \hfill & {2j\mu _1 \xi } \hfill \\
 {2j\mu _1 \xi } \hfill & 0 \hfill \\
\end{array} }} \right),
\]
\[
\beta _{11}^0 =-\left( {{\begin{array}{*{20}c}
 {\lambda _1 +2\mu _1 } \hfill & 0 \hfill \\
 0 \hfill & {-\mu _1 } \hfill \\
\end{array} }} \right)A_1^{-2} \Gamma _1^2 -\left( {{\begin{array}{*{20}c}
 {\lambda _1 \xi ^2} \hfill & 0 \hfill \\
 0 \hfill & {\mu _1 \xi ^2} \hfill \\
\end{array} }} \right),
\]
\[
\gamma _{11}^0 =-\left( {{\begin{array}{*{20}c}
 {\lambda _1 +2\mu _1 } \hfill & 0 \hfill \\
 0 \hfill & {-\mu _1 } \hfill \\
\end{array} }} \right)A_1^{-2} ,
\quad
\delta _{11}^0 =\left( {{\begin{array}{*{20}c}
 0 \hfill & 0 \hfill \\
 0 \hfill & 0 \hfill \\
\end{array} }} \right),
\]
in the conditions of conjugation (\ref{eq6}) we will put
\[
\alpha _{11}^k =\left( {{\begin{array}{*{20}c}
 0 \hfill & {2j\mu _k \xi } \hfill \\
 {2j\mu _k \xi } \hfill & 0 \hfill \\
\end{array} }} \right),
\quad
\beta _{11}^k =-\left( {{\begin{array}{*{20}c}
 {\lambda _k +2\mu _k } \hfill & 0 \hfill \\
 0 \hfill & {-\mu _k } \hfill \\
\end{array} }} \right)A_k^{-2} \Gamma _k^2 -\left( {{\begin{array}{*{20}c}
 {\lambda _k \xi ^2} \hfill & 0 \hfill \\
 0 \hfill & {\mu _k \xi ^2} \hfill \\
\end{array} }} \right),
\]
\[
\gamma _{11}^k =-\left( {{\begin{array}{*{20}c}
 {\lambda _k +2\mu _k } \hfill & 0 \hfill \\
 0 \hfill & {-\mu _k } \hfill \\
\end{array} }} \right)A_k^{-2} ,
\quad
\delta _{11}^k =\left( {{\begin{array}{*{20}c}
 0 \hfill & 0 \hfill \\
 0 \hfill & 0 \hfill \\
\end{array} }} \right),
\]
\[
\alpha _{12}^k =\left( {{\begin{array}{*{20}c}
 0 \hfill & {2j\mu _{k+1} \xi } \hfill \\
 {2j\mu _{k+1} \xi } \hfill & 0 \hfill \\
\end{array} }} \right),
\]
\[
\beta _{12}^k =-\left( {{\begin{array}{*{20}c}
 {\lambda _{k+1} +2\mu _{k+1} } \hfill & 0 \hfill \\
 0 \hfill & {-\mu _{k+1} } \hfill \\
\end{array} }} \right)A_{k+1}^{-2} \Gamma _{k+1}^2 -\left(
{{\begin{array}{*{20}c}
 {\lambda _{k+1} \xi ^2} \hfill & 0 \hfill \\
 0 \hfill & {\mu _{k+1} \xi ^2} \hfill \\
\end{array} }} \right),
\]
\[
\gamma _{12}^k =-\left( {{\begin{array}{*{20}c}
 {\lambda _{k+1} +2\mu _{k+1} } \hfill & 0 \hfill \\
 0 \hfill & {-\mu _{k+1} } \hfill \\
\end{array} }} \right)A_{k+1}^{-2} ,
\quad
\delta _{12}^k =\left( {{\begin{array}{*{20}c}
 0 \hfill & 0 \hfill \\
 0 \hfill & 0 \hfill \\
\end{array} }} \right),
\]
\[
\alpha _{2i}^k =\left( {{\begin{array}{*{20}c}
 1 \hfill & 0 \hfill \\
 0 \hfill & {-1} \hfill \\
\end{array} }} \right),
\quad
\beta _{2i}^k =\left( {{\begin{array}{*{20}c}
 0 \hfill & {j\xi } \hfill \\
 {j\xi } \hfill & 0 \hfill \\
\end{array} }} \right),
\]
\[
\gamma _{2i}^k =\left( {{\begin{array}{*{20}c}
 0 \hfill & 0 \hfill \\
 0 \hfill & 0 \hfill \\
\end{array} }} \right),
\quad
\delta _{2i}^k =\left( {{\begin{array}{*{20}c}
 0 \hfill & 0 \hfill \\
 0 \hfill & 0 \hfill \\
\end{array} }} \right),\quad i=1,2.
\]
Let's apply to a problem (\ref{eq18}), (\ref{eq19}), (23), (24) transforms of Fourier
$F_{n+} $ on the variable $x$. Using identity (\ref{eq17}), we get Cauchy problem
\begin{equation}
\label{eq20}
\frac{d^2}{dt^2}\left( {{\begin{array}{*{20}c}
 {\tilde {\bar {\varphi }}} \hfill \\
 {\tilde {\bar {\psi }}} \hfill \\
\end{array} }} \right)=-c^2\xi ^2\left( {{\begin{array}{*{20}c}
 {\tilde {\bar {\varphi }}} \hfill \\
 {\tilde {\bar {\psi }}} \hfill \\
\end{array} }} \right)-\eta ^2\left( {{\begin{array}{*{20}c}
 {\tilde {\bar {\varphi }}} \hfill \\
 {\tilde {\bar {\psi }}} \hfill \\
\end{array} }} \right)+\left( {{\begin{array}{*{20}c}
 {\bar {p}\left( {\xi ,t} \right)} \hfill \\
 0 \hfill \\
\end{array} }} \right),
\end{equation}
\begin{equation}
\label{eq21}
\left( {{\begin{array}{*{20}c}
 {\tilde {\bar {\varphi }}} \hfill \\
 {\tilde {\bar {\psi }}} \hfill \\
\end{array} }} \right)\left( {\xi ,\eta ,0} \right)=0,\quad
\frac{d}{dt}\left( {{\begin{array}{*{20}c}
 {\tilde {\bar {\varphi }}} \hfill \\
 {\tilde {\bar {\psi }}} \hfill \\
\end{array} }} \right)\left( {\xi ,\eta ,0} \right)=0,
\end{equation}
Here denote
\[
\left( {{\begin{array}{*{20}c}
 {\tilde {\bar {\varphi }}} \hfill \\
 {\tilde {\bar {\psi }}} \hfill \\
\end{array} }} \right)\left( {\eta ,\xi } \right)=F_{n+} \left(
{{\begin{array}{*{20}c}
 {\bar {\varphi }} \hfill \\
 {\bar {\psi }} \hfill \\
\end{array} }} \right)\left( \eta \right),
\]
\[
\left( {{\begin{array}{*{20}c}
 \varphi \hfill \\
 \psi \hfill \\
\end{array} }} \right)
=\sum\limits_{k=1}^n {\theta \left( {l_k -x} \right)\,\theta \left(
{x-l_{k-1} } \right)\,\left( {{\begin{array}{*{20}c}
 {\varphi _k } \hfill \\
 {\psi _k } \hfill \\
\end{array} }} \right)+\theta \left( {x-l_n } \right)\,\left(
{{\begin{array}{*{20}c}
 {\varphi _k } \hfill \\
 {\psi _k } \hfill \\
\end{array} }} \right)} .
\]
Let's result the solution of the problem (\ref{eq20})-(\ref{eq21})
\[
\left( {{\begin{array}{*{20}c}
 {\tilde {\bar {\varphi }}} \hfill \\
 {\tilde {\bar {\psi }}} \hfill \\
\end{array} }} \right)\left( {\eta ,\xi ,t} \right)=\int_0^t {\frac{\sin
\left( {\sqrt {c^2\xi ^2+\eta ^2} \left( {t-\tau } \right)} \right)}{\sqrt
{c^2\xi ^2+\eta ^2} }} \left( {{\begin{array}{*{20}c}
 {\bar {p}\left( {\xi ,\tau } \right)} \hfill \\
 0 \hfill \\
\end{array} }} \right)d\tau .
\]
Let's apply the inverse Fourier transform on $y$ and inverse integral
transform of Fourier series $F_{n+}^{-1} $ on the variable $x$. Using (\ref{eq16}),
we get functions of tension $\varphi _i ,\psi _i $:
\begin{equation}
\label{eq22}
\left( {{\begin{array}{*{20}c}
 {\varphi _i } \hfill \\
 {\psi _i } \hfill \\
\end{array} }} \right)\left( {x,y,t} \right)=\frac{1}{\sqrt {2\pi }
}\int_0^t {\int_{-\infty }^\infty {H_i \left( {x,y-s,t-\tau } \right)} }
\left( {{\begin{array}{*{20}c}
 {p\left( {s,\tau } \right)} \hfill \\
 0 \hfill \\
\end{array} }} \right)dsd\tau ,
\end{equation}
when

$$H_i \left( {x,y-s,t-\tau } \right)=\frac{1}{\sqrt {2\pi } }\int_{-\infty
}^\infty {\left( {-\frac{1}{j\pi }\int_0^\infty {e^{j\xi }u_j } \left(
{x,\eta ,\xi } \right)\cdot \frac{\sin \left( {\sqrt {c^2\xi ^2+\eta ^2}
\left( {t-\tau } \right)} \right)}{\sqrt {c^2\xi ^2+\eta ^2} }d\eta }
\right)} d\xi .$$
The formula (\ref{eq22}) takes the form
\[
\left( {{\begin{array}{*{20}c}
 \varphi \hfill \\
 \psi \hfill \\
\end{array} }} \right)\left( {x,y,t} \right)=
\]
\[
=-\frac{1}{j\sqrt {2\pi } }\int_0^t {\int_{y-c\left( {t-\tau }
\right)}^{y+c\left( {t-\tau } \right)} {H\left( {x,\sqrt {\left( {t-\tau }
\right)^2-\frac{\left( {y-s} \right)^2}{c^2}} } \right)} } \left(
{{\begin{array}{*{20}c}
 {p\left( {s,\tau } \right)} \hfill \\
 0 \hfill \\
\end{array} }} \right)dsd\tau ,
\]
in the case of a homogeneous environment, that is not the dependence of the
$\lambda _i ,\mu _i $-elastic Lama constants of $i$,when $J_0 $ is Bessel
function [3],
\[
H\left( {x,z} \right)=\int_0^\infty {Im\left( {\frac{e^{jx\eta }}{j\eta
\left( {\alpha _{11}^0 +\eta ^2\delta _{11}^0 } \right)+\left( {\beta
_{11}^0 +\eta ^2\gamma _{11}^0 } \right)}} \right)J_0 \left( {\eta z}
\right)d\eta } .
\]
The expressions (\ref{eq22}) for the functions of tension allow to find components
of the vector of displacements $u_i ,v_i ,0$ and the components of the
tension tensor $\sigma _{ix} ,\sigma _{iy} ,\tau _{ixy} $ according to the
formulae (1),(2).

Remark. The dynamic problem of the theory of elasticity for semi space was
considered in the known monograph [15]. However, this problem was solved
without initial conditions. The authors apply the Fourier transform of the
time variable. It leads them to imprecision in the received formulas for the
functions of tension. In our opinion the solution by the method of integral
transforms of Fourier (\ref{eq15}),(\ref{eq16}) on a spatial variable also is more natural.

\section{ Conclusion}

In the work the dynamic problem of elasticity theory are considered: the
problem of oscillations of constructions and buildings, the problem of the
propagation of elastic waves; thermo elastic waves. The method of integral
transforms developed in solving problems. Using the integral transformation
(Fourier, Laplace, Hankel) we came to a more simple task in the pattern
space. Problem of elasticity theory for inhomogeneous bodies studied. These
tasks are of great use in practice. The method of the vector integral
transforms of Fourier with discontinuous coefficients used for the decision
of problems of the theory of elasticity in a piecewise-homogeneous media.
The solution of the dynamic problem in the analytical form found.

\end{document}